\documentclass[english,prd, twocolumn]{revtex4}
\usepackage[T1]{fontenc}
\usepackage[latin9]{inputenc}
\usepackage{textcomp}
\usepackage{amsthm}
\usepackage{amsmath}
\usepackage{amssymb}
\usepackage{esint}

\makeatletter
\@ifundefined{textcolor}{}
{%
 \definecolor{BLACK}{gray}{0}
 \definecolor{WHITE}{gray}{1}
 \definecolor{RED}{rgb}{1,0,0}
 \definecolor{GREEN}{rgb}{0,1,0}
 \definecolor{BLUE}{rgb}{0,0,1}
 \definecolor{CYAN}{cmyk}{1,0,0,0}
 \definecolor{MAGENTA}{cmyk}{0,1,0,0}
 \definecolor{YELLOW}{cmyk}{0,0,1,0}
}
\numberwithin{equation}{section}
\numberwithin{figure}{section}

\usepackage{amsfonts}

\makeatother

\usepackage{babel}
\begin{document}

\title{Pure gauge QCD and holography}

\author{R. C. Trinchero $^{a,b}$ }

\email{ trincher@cab.cnea.gov.ar }

\affiliation{\noindent $^{a}$Instituto Balseiro, Centro Atómico Bariloche, 8400
San Carlos de Bariloche, Argentina.\\
 $^{b}$CONICET, Rivadavia 1917, 1033 Buenos Aires, Argentina.}
\begin{abstract}
Holographic models for the pure gauge QCD vacuum are explored. The
holographic renormalization of these models is considered as required
by a phenomenological approach that takes the $\beta$-functions of
the models as the only input. This approach is done taking the dilaton
as the coordinate orthogonal to the border. This choice greatly simplifies
the analysis and gives a geometrical interpretation for the fixed
points of the renormalization group flow. Examples are constructed
that present asymptotic freedom, confinement of static quarks, either
with vanishing or non-vanishing gluon condensate $G_{2}$. The latter
models require an extension of the dilaton-gravity models already
considered in the literature. This extension is also determined by
the only input, i.e. the $\beta$-function. In addition the restrictions
imposed by the trace anomaly equation are studied. In doing so a holographic
derivation of this equation is presented.
\end{abstract}
\maketitle

\section{Introduction}

The relation between large N gauge theories and string theory \cite{'tHooft:1973jz}
together with the AdS/CFT correspondence \cite{Maldacena:1997re,Gubser:1998bc,Witten:1998qj,malda:1999ti}
have opened new insights into strongly interacting gauge theories.
The application of these ideas to QCD has received significant attention
since those breakthroughs. From the phenomenological point of view,
the so called AdS/QCD approach has produced very interesting results
in spite of the strong assumptions involved in its formulation \cite{Gubser:99,Polch-Strass:00,Sakai:2004cn,Erlich:2005qh,Da_Rold:2005zs,Csaki:2006ji}.
It seems important to further proceed investigating these ideas and
refining the current understanding of a possible QCD gravity dual. 

In this work the phenomenological properties of a 5-dimensional holographic
model for the vacuum of pure gauge QCD are investigated. The interest
in studying that sector of the strong interaction gauge theory stems
from the fact that in spite of its simplicity, it is expected to have
the key features of QCD, namely asymptotic freedom and confinement.
From the holographic point of view, it is interesting to inquire if
it is possible, by an adequate choice of the 5-dimensional action
for background fields, to obtain a reasonable description of both
the IR and UV properties of pure gauge QCD. The model studied in this
work is a simple and phenomenologically motivated extension of 5-dimensional
dilaton gravity already considered in \cite{Gursoy:2007cb,Gursoy:2007er,Gursoy:2010fj}\cite{Goity:2012yj}.
The approach employed in this work take as input the $\beta$-function
of the model and study the resulting phenomenology. In doing so the
holographic renormalization of the theory is considered and adapted
to the phenomenological motivations of this work. The study mentioned
above is greatly simplified by taking the dilaton, or equivalently
the t'Hooft coupling, as the coordinate orthogonal to the border.
This choice is not only useful but also leads to an interesting geometrical
interpretation of the border theory $\beta$-function zeros. 

The features and results of this work are summarized as follows,
\begin{itemize}
\item A simple extension of 5-dimensional dilaton-gravity models is considered
as holographic duals of 4-dimensional pure gauge theories.
\item The holographic renormalization of these models is considered and
adapted to the phenomenological approach. In this approach the $\beta$-functions
of the models is the only input.
\item Taking the dilaton as the radial coordinate greatly simplifies the
phenomenological analysis. In addition it allows to identify borders
of the space with zeros of the $\beta$-function.
\item Models are constructed with vanishing and non-vanishing gluon condensate
$G_{2}$ in the UV. Models with asymptotic freedom and $G_{2}\neq0$
necessarily depart from former dilaton-gravity models and motivate
the extension of these models mentioned above.
\item The restrictions imposed by the trace anomaly equation(TAE) on these
models is studied. In order to understand the meaning of this equation
in holographic terms, it is derived by requiring the independence
of the 5-dimensional action on the renormalization point. In the framework
of the present model such restrictions can be fullfilled and provide
strong constraints on the possible models.
\end{itemize}
The paper is organized as follows. Section II presents the class of
models to be considered and introduces an adequately defined superpotential.
Section III describes the holographic renormalization of these models
 for the case of \textquotedbl{}vacuum\textquotedbl{} metrics. Section
IV deals with the phenomenological approach, the choice of coordinates
mentioned above and the expression in this framework of confinement
and the gluon condensate. In addition, section V presents models with
have vanishing gluon condensate in the UV. This motivates the extension
mentioned above, which is presented in section VI, together with concrete
models constructed out of the perturbative $\beta$-function, which
present confinement and a non-vanishing gluon condensate. Section
VII derives the TAE in holographic language and studies its phenomenological
consequences. Section VIII presents some concluding remarks  and further
interesting issues to be studied.

\section{Action and equations of motion}

The action for the models to be considered is given by,

\begin{eqnarray}
S_{d+1} & = & \frac{N^{2}}{8\pi^{2}}\left(\int_{M_{d+1}}\, d^{d+1}x\;\sqrt{g}\,[-R\right.\nonumber \\
 &  & +V(\phi)+\frac{1}{2}L(\phi)\, g^{ab}\partial_{a}\phi\,\partial_{b}\phi]\nonumber \\
 &  & \left.-2\int_{M_{d}}d^{d}x\,\sqrt{h}\, K\right)\;\;,a,b=0,\cdots,d\label{eq:action}
\end{eqnarray}
it is worth noting that this action for $L(\phi)=1$ reduces to the
one considered in \cite{Gursoy:2007cb,Gursoy:2007er}\cite{Goity:2012yj}.
Bellow, models will be considered with either $L(\phi)=\pm1$ or given
in terms of the $\beta$-function of the border theory. The factor
in front of the parenthesis in (\ref{eq:action}) is the one employed
in \cite{Bianchi-Freedman-Skenderis:01}, where the $N$ refers to
the border theory gauge group, namely $SU(N)$. In this scheme lengths
are measured in units of the metric radius, that in these units is
one%
\footnote{Including adequate powers of the metric radius gives any quantity
in natural units. In this respect it is worth noting that the potential
$V(\phi)$ has dimensions of length to the $-2$. %
}. The equations of motion are,
\begin{eqnarray}
E_{ab}-\frac{1}{2}\, L(\phi)\,\partial_{a}\phi\,\partial_{b}\phi\nonumber \\
+\frac{1}{4}g_{ab}\, L(\phi)(\partial\phi)^{2}+\frac{1}{2}\, g_{ab}V(\phi) & = & 0\label{eq:emunu}\\
\partial_{a}(\sqrt{g}\, g^{ab}\partial_{b}\phi)-\sqrt{g}\;\frac{\partial V(\phi)}{\partial\phi}\nonumber \\
-\frac{1}{2}\frac{\partial L(\phi)}{\partial\phi}\, g^{ab}\partial_{a}\phi\,\partial_{b}\phi & = & 0\label{eq:scalar}
\end{eqnarray}
since the focus is on the vacuum of the boundary field theory, only
metrics and scalar fields having flat boundary space isometry invariance
are considered, thus only solutions for the metric and scalar field
of the following general form are considered, 
\begin{eqnarray}
ds^{2} & = & du^{2}+e^{2A(u)}\,\eta_{\mu\nu}\, dx^{\mu}dx^{\nu}\;,\mu,\nu=0,\cdots,d-1\nonumber \\
\phi & = & \phi(u)\label{eq:global}
\end{eqnarray}
the equations of motion for this choice reduce to,
\begin{eqnarray}
d(d-1)A'^{2}-\frac{L}{2}\phi'^{2}+V(\phi) & = & 0\label{eq:firsteq}\\
A''-\frac{1}{\xi}L\phi'^{2} & = & 0\label{eq:aa}
\end{eqnarray}
\begin{equation}
dLA'\phi'+\left(L\phi'\right)'-\frac{\partial V}{\partial\phi}-\frac{1}{2}\frac{\partial L}{\partial\phi}\phi'^{2}=0\label{eq:restricted-scalar}
\end{equation}
where $\xi=2(1-d)$ and the $'$ denotes derivative respect to $u$.

\subsection{Superpotential}

If a superpotential $W(\phi)$ is defined by,
\begin{equation}
A'=W(\phi)\label{eq:super}
\end{equation}
then,
\[
A''=\phi'\frac{\partial W}{\partial\phi}
\]
comparing with (\ref{eq:aa}) leads to,
\begin{equation}
\phi'=\frac{\xi}{L}\frac{\partial W}{\partial\phi}\label{eq:phi'}
\end{equation}
replacing (\ref{eq:super}) and (\ref{eq:phi'}) in (\ref{eq:firsteq})
leads to,
\begin{equation}
d(d-1)W^{2}-\frac{\xi^{2}}{2L}\left(\frac{\partial W}{\partial\phi}\right)^{2}+V=0\label{eq:first'}
\end{equation}
equation (\ref{eq:restricted-scalar}) in terms of $W$ is,
\begin{eqnarray}
d\xi W\frac{\partial W}{\partial\phi}+\frac{\xi}{L}\frac{\partial W}{\partial\phi}\frac{\partial}{\partial\phi}(\xi\frac{\partial W}{\partial\phi})\nonumber \\
-\frac{1}{2}\frac{\partial L}{\partial\phi}\left(\frac{\xi}{L}\frac{\partial W}{\partial\phi}\right)^{2}-\frac{\partial V}{\partial\phi} & = & 0\label{eq:scal}
\end{eqnarray}
noting that,
\begin{eqnarray*}
\frac{1}{2}\frac{\partial}{\partial\phi}\left(\frac{\xi^{2}}{L}\left(\frac{\partial W}{\partial\phi}\right)^{2}\right) & = & \frac{\xi}{L}\frac{\partial W}{\partial\phi}\frac{\partial}{\partial\phi}(\xi\frac{\partial W}{\partial\phi})\\
 &  & -\frac{1}{2}\frac{\partial L}{\partial\phi}\left(\frac{\xi}{L}\frac{\partial W}{\partial\phi}\right)^{2}
\end{eqnarray*}
shows that (\ref{eq:scal}) is the same as,
\[
\frac{\partial}{\partial\phi}\left[\frac{d\xi}{2}W^{2}-V+\frac{\xi^{2}}{2L}\left(\frac{\partial W}{\partial\phi}\right)^{2}\right]=0
\]
thus (\ref{eq:first'}) implies (\ref{eq:scal}), if everything is
described in terms of the superpotential $W$ defined by equations
(\ref{eq:super}) and (\ref{eq:phi'}).

\subsection{The on-shell action}

In an analogous way as in \cite{Goity:2012yj} the on-shell $d+1$-dimensional
action can be computed and is given by,
\begin{equation}
S_{d+1}=\frac{N^{2}}{8\pi^{2}}\xi W(\phi)\sqrt{g}\;\;,\sqrt{g}=e^{dA}\label{eq:sonshell}
\end{equation}
The energy-momentum tensor is obtained recalling that, according to
the correspondence, 
\begin{equation}
S=\int_{M_{d}}d^{d}x\;\sqrt{g}\, g_{\mu\nu}T^{\mu\nu},
\end{equation}
leading to,
\begin{equation}
T_{\mu}^{\mu}=\frac{2}{\sqrt{g}}g_{\mu\nu}\frac{\delta S_{d+1}}{\delta g_{\mu\nu}^{R}}=-d\frac{N^{2}}{4\pi^{2}}\xi W(u)\label{eq:tmumu}
\end{equation}
this result coincides with one obtained from the expression of this
trace in terms of the extrinsic curvature\cite{Brown-York:93}\cite{Bala-Krauss:99}\cite{Myers:1999psa}
on the border of this space.

\section{Holographic Renormalization\label{sec:Renormalization}}

A basic idea in the holographic renormalization group is the relation
between the radial coordinate(orthogonal to the border theory coordinates)
and the energy scale at which the border theory is observed. This
relation is given by, 
\begin{equation}
\mu=e^{2A(u)}\label{eq:scale}
\end{equation}
and this last factor has the information about how the scale of energies
depends on the radial coordinate. This is a direct consequence of
the form (\ref{eq:global}) of the metric in these coordinates.

Replacing the solutions of the equations of motion for $A(u)$ and
$\phi(u)$ in $S_{d+1}$ leads to a divergent expression when $u\to\infty$,
the UV border. This expression can be regularized\cite{Witten:1998qj}
by introducing a cut-off $u_{0}$ in the coordinate $u$. Next a substraction
procedure is employed. It is important that this procedure does not
spoil the symmetries of the theories under consideration. For example
it should be such that in the AdS case it leads to a expression for
the renormalized on-shell five dimensional action such that it implies
a vanishing trace of the border theory energy-momentum tensor. That
is, in the AdS case, conformal symmetry should be maintained by the
renormalization procedure. This can be achieved by considering the
leading asymptotic behavior of the solutions and defining the renormalized
action $S_{d+1}^{R}$ by subtracting to the five dimensional on-shell
bare action, the counterterm action $S_{d+1}^{C}$ defined by evaluating
the bare action with these asymptotic expressions and with the same
cut-off $u_{0}$ as above . This procedure was proposed in \cite{hawking-horowitz:1996}
and applied to holographic models in \cite{Bala-Krauss:99,Myers:1999psa},
it leads to a finite expression for the substracted action and it
preserves conformal symmetry in the AdS case. Thus,
\[
S_{d+1}^{R}=S_{d+1}-S_{d+1}^{C}
\]

As mentioned at the end of the last section the trace of the energy-momentum
tensor derived from the on-shell bare gravitational action is given
by,
\[
T_{\mu}^{\mu}=\frac{2}{\sqrt{g}}g_{\mu\nu}\frac{\delta S_{d+1}}{\delta g_{\mu\nu}^{R}}=-d\frac{N^{2}}{4\pi^{2}}\xi W(u)
\]
Based on this bare relation, the renormalized trace of the energy
momentum tensor $^{R}T_{\mu}^{\mu}$ and the renormalized superpotential
$W^{R}$ are defined by,
\[
\,^{R}T_{\mu}^{\mu}=\frac{2}{\sqrt{g}}g_{\mu\nu}\frac{\delta S_{d+1}^{R}}{\delta g_{\mu\nu}^{R}}=-d\frac{N^{2}}{4\pi^{2}}\xi W^{R}(u)
\]
given the renormalized superpotential, other quantities are defined
by maintaining the bare relations. For example the renormalized wrap
factor can be obtained by integrating the equation,
\[
\frac{dA^{R}}{du}=W^{R}(u)\Rightarrow A^{R}(u)=C_{A}+\int du\, W^{R}(u)
\]
the integration constant $C_{A}$ is fixed by giving its value for
a given point $u_{R}$ of the radial coordinate %
\footnote{This corresponds to choosing one of the classical solutions of the
equations of motion that differ from one another by the value of this
constant\cite{boer-verlinde:2000,Fukuma:00}.%
}. This value $u_{R}$ plays the role of the renormalization scale
or substraction point in field theory renormalization. As usual physical
results should be independent of $u_{R}$. As will be shown below,
for the case of confining solutions, the constant $C_{A}$ is related
to the string tension of the linear potential between static quarks.

The phenomenological approach employed in this work is to take the
$\beta$-function as input of the model. The beta function of the
gauge theory can be obtained from the AdS/CFT identification of the
$SU(N)$ Yang-Mills renormalized coupling $g_{YM}$ with the dilaton
profile according to \cite{malda:1999ti}:
\begin{equation}
\frac{\lambda}{N}=\frac{g_{YM}^{2}}{4\pi}=e^{\phi}\label{eq:lambda}
\end{equation}
where $\lambda$ is the renormalized t'Hooft coupling. Using also
(\ref{eq:scale}) the beta function is given by,
\begin{equation}
\beta(\lambda)=\frac{d\lambda}{d\log\mu}=\frac{d\lambda}{dA^{R}}\label{eq:beta}
\end{equation}
The models considered in this work have $L(\phi)$ equal to $\pm1$
or given in terms of the $\beta$-function, thus $L(\phi),$ as the
$\beta$-function itself, is subject to no renormalization. Indeed,
in the phenomenological approach given bellow they are considered
as external input. From now on all indices $R$ indicating renormalized
quantities will not be included, every quantity appearing in the following
equations is a renormalized one, unless otherwise is stated.

\section{Phenomenological approach and choice of coordinates}

As mentioned above the models considered bellow take as input the
$\beta$-function depending on the t'Hooft coupling $\lambda$ defined
in (\ref{eq:lambda}). From this input the exponent $A$ appearing
in the warp factor that defines the metric (\ref{eq:global}) can
be obtained as a function of $\lambda$. Indeed, the second equality
in (\ref{eq:beta}) implies that,
\begin{equation}
A(\lambda)=c_{A}+\int\frac{d\lambda}{\beta(\lambda)}\label{eq:abeta}
\end{equation}
furthermore the superpotential $W$ can be obtained as a function
of $\lambda$ as follows,
\begin{eqnarray}
W & = & A'=\frac{dA}{d\phi}\phi'=\frac{\lambda\xi}{\beta(\lambda)L}\frac{\partial W}{\partial\phi}\nonumber \\
 & \Downarrow\nonumber \\
\ln W & = & c_{W}+\int d\lambda\frac{L(\lambda)\beta(\lambda)}{\xi\lambda^{2}}\label{eq:Wl}
\end{eqnarray}
where in the third equality (\ref{eq:phi'}) was employed. Thus knowing
the beta function as a function of $\lambda$ determines the superpotential
and the solution of the equations of motion for the warp factor as
a function of $\lambda$. 

The physical properties of the models considered in this work, in
the domain wall coordinates employed in (\ref{eq:global}), require
the computation of both $A(u)$ and $\phi(u)$. The dependence of
the dilaton on $u$ can in principle be obtained from (\ref{eq:beta})
and (\ref{eq:Wl}),
\begin{equation}
\phi'=\frac{\beta(\lambda)W(\lambda)}{\lambda}\Rightarrow u=C+\int\frac{d\lambda}{\beta(\lambda)W(\lambda)}\label{eq:phip}
\end{equation}
and obtaining the inverse function of the r.h.s in the last equation.
Therefore the obtention of a closed analytical expression for $\phi(u)$
is not always possible in general. However there is an alternative
approach which is quite interesting in itself and permits to obtain
the physical results without having to deal with inversions of the
sort mentioned above. This approach consists in taking the dilaton
as the radial coordinate, indeed the metric (\ref{eq:global}) can
be written in terms of this coordinate as follows,

\begin{eqnarray}
ds^{2} & = & du^{2}+e^{2A(u)}\,\eta_{ij}\, dx^{i}dx^{j}\nonumber \\
 & = & \frac{d\phi^{2}}{\phi'(\phi)^{2}}+e^{2A(\phi)}\eta_{\mu\nu}dx^{\mu}dx^{\nu}\nonumber \\
 & = & \frac{d\lambda^{2}}{\beta(\lambda)^{2}W(\lambda)^{2}}+e^{2A(\lambda)}\eta_{\mu\nu}dx^{\mu}dx^{\nu}\label{eq:lambda-metric}
\end{eqnarray}
where the last equality follows form (\ref{eq:phip}). Recalling that
the borders of the space are given by the singularities and zeros
of the coefficients in the metric, it becomes clear that \emph{the
zeros of the beta function correspond to borders of the space}. This
last statement is very interesting because it gives a clear geometrical
interpretation of an important field theoretical quantity. It should
be remarked that its derivation is very simple, quite general and
not restricted to the models under consideration. Indeed only the
identification (\ref{eq:lambda}) and the consideration of global
metrics (\ref{eq:global}), that correspond to studying the vacuum
of the border field theory, are employed to derive this interpretation. 

Next, the condition for confinement and the computation of the gluon
condensate are considered in this scheme.

\subsection{Confinement}

This is studied by computing the static quark potential through the
consideration of a rectangular Wilson loop. This is done using the
Nambu-Goto action as in \cite{Maldacena:1998im}. For the case of
a non-AdS metric this study is considered in \cite{kinar-sonnenschein:1998}
where a condition for confinement is given in terms of the metric
tensor components. For the metric in (\ref{eq:lambda-metric}) the
following two functions are considered in \cite{kinar-sonnenschein:1998}
,
\[
f(\lambda)=e^{2A_{S}(\lambda)}\;\;,\;\; g(\lambda)=\frac{e^{2A_{S}(\lambda)}}{\beta(\lambda)W(\lambda)}
\]
where $A_{S}(\lambda)$ denotes the string frame warp exponent given
by,
\[
A_{S}(\lambda)=A(\lambda)+\frac{2}{d-1}\phi(\lambda)
\]
The condition states that there is a linear potential between static
quarks if $f(\lambda)$ has a minimum $\lambda_{0}$ and the string
tension is given by $f(\lambda_{0})$. According to (\ref{eq:abeta})
the function $f(\lambda)$ is given in terms of the $\beta$ function
by,
\[
f(\lambda)=e^{2\left(c_{A}+\int\frac{d\lambda}{\beta(\lambda)}+\frac{2}{d-1}\ln(\frac{\lambda}{N})\right)}
\]
the condition for having an extremum being,
\[
\frac{df(\lambda)}{d\lambda}=0=e^{2\left(c_{A}+\int\frac{d\lambda}{\beta(\lambda)}+\frac{2}{d-1}\ln(\frac{\lambda}{N})\right)}\left(\frac{1}{\beta(\lambda)}+\frac{2}{(d-1)\lambda}\right)
\]
thus for a non-vanishing string tension the condition for having a
extremum at $\lambda_{0}$ is,
\begin{equation}
\beta(\lambda_{0})=-\frac{(d-1)}{2}\lambda_{0}\label{eq:conf-con}
\end{equation}
with the string tension given by,
\[
\sigma=e^{2\left(c_{A}+\int\frac{d\lambda}{\beta(\lambda)}+\frac{2}{d-1}\ln(\frac{\lambda}{N})\right)}|_{\lambda=\lambda_{0}}
\]
this relation is remarkable in the sense that it gives a condition
for confinement and a expression for the string tension, which make
no reference to its higher dimensional holographic origin%
\footnote{In ref. \cite{Gursoy:2007er}, a condition for confinement is given
in terms of the IR asymptotics of the beta function, this is done
for geometries that are asymptotically AdS. The condition given in
(\ref{eq:conf-con}) reduces to the one in \cite{Gursoy:2007er} for
these cases. The point in considering this extension is that as argued
in \cite{Goity:2012yj} the matching with the perturbative QCD $\beta$-function
in the UV leads, in the framework of these models, to a metric that
is not asymptotically AdS. More precisely, the leading term is AdS(proportional
to $u$ in domain wall coordinates) but there are additional terms(logarithmic
in $u$) which are not AdS and diverge when approaching the UV border
($u\to\infty$).%
}.

\subsection{The gluon condensate}

Among important fundamental non-perturbative effects in QCD, the existence
of a non-vanishing gluon condensate $G_{2}$ was early on identified
\cite{Shifman1979385}. It has important manifestations in hadron
phenomenology, and there are indications of its non-vanishing from
lattice QCD \cite{DiGiacomo:1981wt,Ilgenfritz:2012nca}. However it
should be mentioned that it is not completely obvious from the data
that it should be non-vanishing\cite{Holdom:08}.

The gluon condensate $G_{2}$ is proportional to the v.e.v. of $Tr[F^{2}]$,
i.e.%
\footnote{The field strength tensor and pure gauge theory action employed in
(\ref{eq:g2}) are given by,
\[
F_{\mu\nu}=\partial_{\mu}A_{\nu}-\partial_{\nu}A_{\mu}+[A_{\mu},A_{\nu}]\;\;,\mathcal{L=}\frac{1}{2g^{2}}Tr[F^{2}]
\]
in terms of the rescaled field $A_{\mu}^{a}=g\tilde{A}_{\mu}^{a}$,
the field strength, Lagrangian density and gluon condensate are given
by,
\[
\tilde{F}_{\mu\nu}^{a}=\partial_{\mu}\tilde{A}_{\nu}^{a}-\partial_{\nu}\tilde{A}_{\mu}^{a}+gf^{abc}[\tilde{A}_{\mu}^{b},\tilde{A}_{\nu}^{c}]\;\;,\mathcal{L=}\frac{1}{2}Tr[\tilde{F}^{2}]
\]
\[
G_{2}=\frac{-g^{2}}{2\pi^{2}}<Tr[\tilde{F}^{2}]>=-\frac{1}{2\pi^{2}}<Tr[F^{2}]>
\]
thus this definition of $G_{2}$ is the same as the one in (\cite{DiGiacomo:1981wt}). %
},
\begin{equation}
G_{2}=\frac{-1}{2\pi\text{\texttwosuperior}}<Tr[F^{2}]>\label{eq:g2}
\end{equation}
this quantity is an approximate renormalization group invariant in
the UV. Indeed the trace anomaly equation%
\footnote{This relation is at the basis of the gluon condensate lattice computations\cite{DiGiacomo:1981wt}.%
},
\[
T_{\mu}^{\mu}=-\frac{N}{8\pi}\frac{\beta(\lambda)}{\lambda^{2}}Tr[F^{2}]=\frac{\pi N}{4}\frac{\beta(\lambda)}{\lambda^{2}}\, G_{2}
\]
shows that in the UV, where the $\beta$-function is well approximated
by $\beta(\lambda)=-b_{0}\lambda^{2}$, the gluon condensate is up
to a constant the same as the trace of the energy momentum tensor.
From the holographic point of view $G_{2}$ can be computed using
the following source-operator correspondence,
\[
\frac{1}{g_{YM}^{2}}=\frac{e^{-\phi}}{4\pi}\rightsquigarrow\frac{1}{2}Tr[F^{2}]
\]
leading to
\begin{equation}
\frac{2}{\sqrt{g}}\frac{\delta S_{d+1}^{R}}{\delta\left(\frac{1}{g_{YM}^{2}}\right)}=<Tr[F^{2}]>=-4\pi\frac{\lambda^{2}}{N}\left(\frac{2}{\sqrt{g}}\frac{\delta S_{d+1}^{R}}{\delta\lambda}\right)\label{eq:corresp}
\end{equation}
thus giving the following holographic expression for $G_{2}$,
\begin{equation}
G_{2}=\frac{2}{\pi}\frac{\lambda^{2}}{N}\left(\frac{2}{\sqrt{g}}\frac{\delta S_{d+1}^{R}}{\delta\lambda}\right)=\frac{\lambda^{2}N}{2\pi^{3}}\xi\frac{dW(\lambda)}{d\lambda}\label{eq:gluon condesate}
\end{equation}
where in the last equality the renormalized version of (\ref{eq:sonshell})
was employed.

\section{Examples with $L=\pm1$}

In this section two sample models are considered. On of them, referred
to as non-perturbative, has been considered in \cite{Goity:2012yj}
were it was chosen so as to be able to analytically perform the inversion
mentioned in the last section. This inversion should be implemented
for example to obtain $\phi(u)$ from (\ref{eq:phip}). Thus in this
example it is possible to check explicitly that the physical results
regarding confinement and the gluon condensate are the same using
the domain wall coordinate $u$ or the t'Hooft coupling as a coordinate.
In addition this example is asymptotically AdS. On the perturbative
side, the two loop perturbative beta function is considered. For this
example the inversions mentioned above can not be performed analytically,
however all the physical information that is obtained for the first
example can be also obtained in this case thanks to the choice of
coordinates appearing in the previous section

\subsection{Non-perturbative }

The beta function of this example is given by\cite{Goity:2012yj},
\[
\beta(\lambda)=-\frac{\alpha\,\lambda\ln\frac{\lambda}{N}}{1-\frac{\alpha}{2\xi}\ln^{2}\frac{\lambda}{N}}
\]
this implies, using (\ref{eq:abeta}) and (\ref{eq:Wl}), that $A(\lambda)$
and $W(\lambda)$ are given in this case by,
\begin{eqnarray}
A(\lambda) & = & C_{A}-\frac{1}{24}\ln^{2}\frac{\lambda}{N}-\frac{1}{\alpha}\ln\left(\log\frac{\lambda}{N}\right)]\nonumber \\
W(\lambda) & = & e^{C_{W}}\left(12+\alpha\ln^{2}\frac{\lambda}{N}\right)\label{eq:wlnonper}
\end{eqnarray}
this $\beta$-function has a zero at $\lambda=N$ where $W(\lambda)$
is finite, thus indicating that there is a border of the space at
that value. This border is an ultraviolet one since the energy scale
$\mu=e^{2A}$ diverges at $\lambda=N$. On the other hand for $\lambda\to\infty$
, the $\beta$-function diverges and the energy scale goes to zero
signaling an infrared border. Having vanishing Yang-Mills coupling
in the UV indicates the identification,
\begin{equation}
\frac{\lambda}{N}-1=\frac{g_{YM}^{2}}{4\pi}\label{eq:lambda-np}
\end{equation}
this is different form (\ref{eq:lambda}) and produces a change in
the expression (\ref{eq:gluon condesate}) which for the identification
(\ref{eq:lambda-np}) is,
\[
G_{2}=\frac{2N}{\pi}(\frac{\lambda}{N}-1)^{2}\left(\frac{2}{\sqrt{g}}\frac{\delta S_{d+1}^{R}}{\delta\lambda}\right)
\]

\subsubsection{Confinement}

Condition (\ref{eq:conf-con}) in this case leads to,
\[
-\frac{\alpha\,\lambda_{0}\log\frac{\lambda_{0}}{N}}{1-\frac{\alpha}{2\xi}\log^{2}\frac{\lambda_{0}}{N}}=-\frac{(d-1)}{2}\lambda_{0}
\]
whose solutions for $d=4$ and $N=3$ are ,
\[
\lambda_{0}^{\pm}=e^{\frac{4\alpha\pm2\sqrt{-3\alpha+4\alpha^{2}}+\alpha\text{Log}[3]}{\alpha}}
\]
the - sign corresponding to a minimum of $A_{s}(\lambda)$ the + to
a maximum . Thus the string tension is given by ,
\begin{eqnarray*}
\sigma & = & e^{2A_{S}(\lambda_{0}^{-})}\\
 & = & e^{2C_{A}-\frac{1}{12}\text{Log}\left[e^{4-\frac{2\sqrt{\alpha(-3+4\alpha)}}{\alpha}}\right]^{2}}\times\\
 &  & \left(e^{4-\frac{2\sqrt{\alpha(-3+4\alpha)}}{\alpha}}\right)^{4/3}\left(4-\frac{2\sqrt{\alpha(-3+4\alpha)}}{\alpha}\right)^{-\frac{2}{\alpha}}
\end{eqnarray*}
for $\alpha=1$ these expressions reduce to,
\[
\lambda_{0}^{-}|_{\alpha=1}=3e^{2}\;\;,\sigma|_{\alpha=1}=\frac{1}{4}e^{\frac{7}{3}+2C_{A}},
\]
The same calculation can be done in terms of the coordinate $u$ obtaining
the same results. The connection between both being given explicitly
by the change of coordinates\cite{Goity:2012yj},
\[
\lambda=Ne^{\phi(u)}=Ne^{C\, e^{-\alpha\; u}}
\]
with the constant $C_{A}$ related to $C$ by $C_{A}=\frac{\ln C}{\alpha}$.
This relation shows that the ultraviolet $\lambda\to N$ corresponds
for $\alpha>0$ to $u\to\infty$ and the infrared $\lambda\to\infty$
to $u\to-\infty$ and interchanged if $\alpha<0$.

\subsubsection{Gluon condensate}

Using (\ref{eq:sonshell}) and (\ref{eq:gluon condesate}), the following
expression for the gluon condensate is obtained,
\begin{eqnarray*}
G_{2}(\lambda) & = & \frac{(\lambda-N)^{2}N}{2\pi^{3}}\xi\frac{\partial W(\lambda)}{\partial\lambda}\\
 & = & \frac{e^{C_{W}}\alpha N\xi}{\pi^{3}}\frac{(\lambda-N)^{2}}{\lambda}\ln(\frac{\lambda}{N})
\end{eqnarray*}
whose power series expansion in the UV, i.e. for $\lambda\to N$ is,
\begin{eqnarray*}
G_{2}(\lambda) & = & \frac{e^{C_{W}}\alpha N\xi}{\pi^{3}}\left(\frac{(\lambda-N)^{3}}{N^{2}}-\frac{(\lambda-N)^{4}}{2N^{3}}+\cdots\right)\\
 & = & \frac{e^{C_{W}}\alpha N^{2}\xi}{\pi^{3}}\left(\left(\frac{g_{YM}^{2}}{4\pi}\right)^{3}-\frac{1}{2}\left(\frac{g_{YM}^{2}}{4\pi}\right)^{4}+\cdots\right)
\end{eqnarray*}
showing that in this case $G_{2}$ goes to zero in the UV.

\subsection{Perturbative }

The 2-loop perturbative $\beta$-function is considered,
\[
\beta_{2l}(\lambda)=-\frac{11}{6\pi}\lambda^{2}-\frac{17}{12\pi^{2}}\lambda^{3}
\]
using (\ref{eq:abeta}) and (\ref{eq:Wl}), $A(\lambda)$ and $W(\lambda)$
are obtained,
\begin{eqnarray*}
A(\lambda) & = & C_{A}-\frac{12}{22}\pi^{2}\left[-\frac{1}{\pi\lambda}\right.\\
 &  & \left.-\frac{17}{22\pi^{2}}\left(\text{Log}[\lambda]\text{-Log}[22\pi+17\lambda]\right)\right]
\end{eqnarray*}

\begin{equation}
W(\lambda)=\exp\left(C_{W}-\frac{\lambda(44\pi+17\lambda)}{24\pi^{2}\xi}\right)\label{eq:wpert}
\end{equation}
the energy scale is given by,
\[
\mu=e^{A(\lambda)}=e^{C_{A}+\frac{6\pi}{11\lambda}}\left(\frac{\lambda}{e^{C_{A}+\frac{6\pi}{11\lambda}}}\right)^{51/121}
\]
which shows that $\lambda\to0$ corresponds to the UV and that there
is a minimum energy scale to be explored by this model given by,
\[
\lim_{\lambda\to\infty}\mu=\frac{e^{C_{A}}}{17^{51/121}}
\]

\subsubsection{Confinement}

Condition (\ref{eq:conf-con}) in this case leads to a cubic equation,
\[
\frac{(d-1)}{2}\lambda_{0}-\frac{11}{6\pi}\lambda_{0}^{2}-\frac{17}{12\pi^{2}}\lambda_{0}^{3}=0
\]
the solution that corresponds to a minimum of $A_{S}(\lambda)$ is,
\[
\lambda_{0}=\frac{1}{17}\left(-11+\sqrt{19+102d}\right)\pi\underset{_{d=4}}{=}1.785
\]
with the string tension given by,
\[
\sigma=e^{2A_{S}(\lambda_{0})}\underset{_{d=4}}{=}0.27\, e^{0.96\,+C_{A}}\left(\frac{1}{N}\right)^{2/3}
\]

\subsubsection{The gluon condensate}

Using (\ref{eq:gluon condesate}) and (\ref{eq:wpert}) leads to,
\begin{eqnarray*}
G_{2} & = & \frac{\lambda^{2}N}{2\pi^{3}}\xi\frac{\partial W(\lambda)}{\partial\lambda}\\
 & =-\frac{\lambda^{2}N}{2\pi^{3}} & \frac{e^{\text{Cw}-\frac{\lambda(44\pi+17\lambda)}{24\pi^{2}\xi}}(22\pi+17\lambda)}{12\pi^{2}}
\end{eqnarray*}
whose series expansion around $\lambda\to0$ is,
\[
G_{2}=-\frac{11\left(e^{C_{W}}N\right)\lambda^{2}}{12\pi^{4}}-\frac{\left(e^{C_{W}}N(-121+51\xi)\right)\lambda^{3}}{72\left(\pi^{5}\xi\right)}
\]
which shows that also in this case the gluon condensate vanishes in
the UV($\lambda\to0$).

\section{Non-vanishing gluon condensate and $L(\lambda)$}

The superpotential $W(\lambda)$ can be expressed in terms of the
gluon condensate $G_{2}(\lambda)$ using (\ref{eq:gluon condesate}),
\begin{equation}
W(\lambda)=\tilde{C}_{w}+\frac{2\pi^{3}}{\xi N}\int d\lambda\,\frac{G_{2}(\lambda)}{\lambda^{2}}\label{eq:ws}
\end{equation}
In addition the relation (\ref{eq:Wl}) allows to express the $\beta$-function
in term s of the superpotential%
\footnote{Eq. (\ref{eq:beta-ws}) shows that given a $\beta$-function $W$
is determined up to a multiplicative constant, $e^{C_{w}}$. %
},
\begin{equation}
\beta=\frac{\xi\lambda^{2}}{LW}\frac{dW}{d\lambda}\label{eq:beta-ws}
\end{equation}
using (\ref{eq:ws}) the $\beta$-function is given in terms of $G_{2}$
by
\[
\beta=\frac{G_{2}(\lambda)}{L(\lambda)\left(\tilde{C}_{w}+\frac{2\pi^{3}}{\xi N}\left(\int d\lambda\,\frac{G_{2}(\lambda)}{\lambda^{2}}\right)\right)}\frac{2\pi^{3}}{N}
\]
This equation shows that if a theory has $L=\pm1$ and a non-vanishing
gluon condensate in the UV, then the beta function near $\lambda=0$
grows linearly with $\lambda$ with positive(negative) coefficient.
Indeed taking $G_{2}(\lambda)$ to be a constant $g_{2}$ leads to,
\[
\beta_{g_{2}}^{L=\pm1}=\pm\frac{g_{2}}{\tilde{C}_{w}-\frac{2\pi^{3}g_{2}}{\xi N\lambda}}\underset{_{\lambda\to0}}{\simeq}\mp\frac{\xi N\lambda}{2\pi^{3}}\underset{_{d=4}}{=}\pm\frac{6N\lambda}{2\pi^{3}}
\]
Next it is shown that it is possible to choose expressions for $L(\lambda)$
solely determined by the $\beta$-function that lead to a non-vanishing
gluon condensate in the UV, without restricting the choice of the
$\beta$-function. Two choices will be considered.
\begin{enumerate}
\item $L(\lambda)=\frac{-\xi\lambda}{\beta(\lambda)}$ . This choice leads
to, 
\[
W(\lambda)=\exp\left(C_{W}-\int\frac{d\lambda}{\lambda}\right)=\frac{e^{C_{W}}}{\lambda}
\]
which gives,
\[
G_{2}=-\frac{e^{C_{W}}N\xi}{2\pi^{3}}\underset{_{d=4}}{=}\frac{6\, e^{C_{W}}N}{2\pi^{3}}
\]
it is noteworthy that the constant gluon condensate determines the
integration constant $C_{W}$.
\item $L(\lambda)=\frac{-\xi\lambda}{\beta_{as}(\lambda)}$. Where $\beta_{as}(\lambda)$
denotes the asymptotic expression of $\beta(\lambda)$ for $\lambda\to0$.
This gives,
\[
W(\lambda)=\exp\left(C_{W}-\int\frac{d\lambda}{\lambda}\frac{\beta(\lambda)}{\beta_{as}(\lambda)}\right)
\]
by definition of $\beta_{as}(\lambda)$ it holds that,
\[
\frac{\beta(\lambda)}{\beta_{as}(\lambda)}=1+f(\lambda)\;\;,\lim_{\lambda\to0}f(\lambda)=0
\]
 leading to the following expression for $G_{2}$,
\[
G_{2}(\lambda)=\frac{-e^{C_{W}}N\xi}{2\pi^{3}}[1+f(\lambda)]\exp\left(-\int d\lambda f(\lambda)\right)
\]
which in the UV has a constant value given, as in the previous case,
by,
\[
\lim_{\lambda\to0}G_{2}(\lambda)=\frac{-e^{C_{W}}N\xi}{2\pi^{3}}
\]

\end{enumerate}

\section{Holography and the trace anomaly equation}

\subsection{Derivation of the TAE}

In this section it is shown that the TAE can be obtained, in the holographic
context, by requiring the independence on the renormalization scale
$u_{R}$ of the renormalized on-shell five dimensional action as a
function of its boundary values. This  requirement is analogous to
the one employed \cite{Osborn:91} for deriving the TAE solely in
terms of the border field theory.

This requirement is,
\[
\frac{dS_{d+1}(A,\phi)}{du_{R}}=0\Rightarrow\frac{\delta S_{d+1}}{\delta A}A'+\frac{\delta S_{d+1}^{R}}{\delta\phi}\phi'=0
\]
which, recalling that,
\begin{eqnarray*}
-\frac{2}{\sqrt{g}}g_{\mu\nu}\frac{\delta\sqrt{g}}{\delta g_{\mu\nu}} & = & \frac{\delta\sqrt{g}}{\delta A}\;\;,\sqrt{g}=e^{dA}
\end{eqnarray*}
and that (\ref{eq:lambda}) implies,
\[
\frac{\delta S_{d+1}}{\delta\phi}=-\frac{1}{g_{YM}^{2}}\frac{\delta S_{d+1}}{\delta(1/g_{YM}^{2})}
\]
leads to,
\[
-\frac{2}{\sqrt{g}}g_{\mu\nu}\frac{\delta S_{d+1}}{\delta g_{\mu\nu}}-\frac{\phi'}{A'}\frac{1}{g_{YM}^{2}}\left(\frac{2}{\sqrt{g}}\frac{\delta S_{d+1}}{\delta(1/g_{YM}^{2})}\right)=0
\]
recalling the holographic version of the border theory energy-momentum
tensor $T_{\mu\nu}$ vacuum expectation value (\ref{eq:tmumu}) ,
the definition of the $\beta$-function (\ref{eq:beta}), which implies
$\beta(g_{YM}^{2})=g_{YM}^{2}\,\frac{d\phi}{dA}=g_{YM}^{2}\,\frac{\phi'}{A'}$
and the correspondence (\ref{eq:corresp}) leads to,
\begin{equation}
<T_{\mu}^{\mu}>=-\frac{\beta(g_{YM}^{2})}{g_{YM}^{4}}<Tr[F^{2}]>\label{eq:tae1}
\end{equation}
which is the TAE.

\subsection{Phenomenology and TAE }

Replacing (\ref{eq:tmumu}), and using (\ref{eq:sonshell}) to compute
the derivative in the r.h.s. of (\ref{eq:tae1}) leads to,
\begin{equation}
W(\lambda)=-\frac{1}{d}\beta(\lambda)\frac{dW(\lambda)}{d\lambda}\label{eq:taew}
\end{equation}
comparing this equation with (\ref{eq:Wl}) leads an expression of
$L(\lambda)$ in terms of the $\beta$-function,
\begin{equation}
L(\lambda)=-\frac{d\xi\lambda^{2}}{\beta(\lambda)^{2}}\label{eq:llam}
\end{equation}
the solution of (\ref{eq:taew}) for $W(\lambda)$ is,
\begin{equation}
W=e^{C_{W}-d\int\frac{d\lambda}{\beta(\lambda)}}\label{eq:wbeta}
\end{equation}
 replacing (\ref{eq:taew}) and (\ref{eq:wbeta}) in (\ref{eq:gluon condesate})
implies,
\begin{equation}
G_{2}=-d\xi\frac{\lambda{}^{2}N}{2\pi^{3}}\frac{e^{C_{W}-d\int\frac{d\lambda}{\beta(\lambda)}}}{\beta(\lambda)}\label{eq:g2beta}
\end{equation}
Taking $\beta(\lambda)=d\lambda$ leads to a non-vanishing UV condensate
given by,
\[
G_{2}=-\xi\frac{N}{2\pi^{3}}e^{C_{W}}\underset{_{d=4}}{=}\frac{18}{2\pi^{3}}e^{C_{W}}
\]
However, taking for $\beta(\lambda)$ the 1-loop result $-\frac{11}{6\pi}\lambda^{2}$
leads to,
\[
G_{2}=d\frac{N}{2\pi^{3}}\xi\left(\frac{6\pi}{11}\right)e^{C_{W}-\frac{6\pi d}{11\lambda}}
\]
which leads to $G_{2}=0$ in the UV. 

This results indicate that in the framework of the models considered
in this work and described by the $d+1$-dimensional action given
by (\ref{eq:action}), the requirements imposed by the TAE can be
met. They restrict $L(\phi)$ to fulfill (\ref{eq:llam}). The case
of vanishing gluon condensate $G_{2}=0$, corresponds to $\beta$-functions
such that,
\[
\lim_{\lambda\to0}\frac{e^{-d\int\frac{d\lambda}{\beta(\lambda)}}}{\beta(\lambda)}=0
\]
and is fulfilled by the perturbative $\beta$-functions. For $G_{2}\neq$0
 the $\beta$-function should include a term linear in $\lambda$
with positive coefficient.

\section{Conclusions and outlook}

In this work the description of the pure gauge QCD vacuum by means
of a 5-dimensional two derivative dilaton gravity like holographic
theory has been considered. This has been done taking as only input
the $\beta$-function of the border gauge field theory. It has been
shown that models can be constructed with the following properties: 
\begin{enumerate}
\item Asymptotic freedom and UV behavior matching the perturbative $\beta$-function.
\item Confinement of static quarks. 
\end{enumerate}
Regarding the gluon condensate $G_{2}$ it has been shown that in
order to have $G_{2}\neq0$ in models with property 1. and 2., an
extension of the two derivative dilaton gravity model considered in
the literature should be included. In this work this is done by including
a dilaton dependent factor in front of the dilaton kinetic term, which
is either a constant or given in terms of the $\beta$-function. Furthermore
the validity or not of the trace anomaly equation(TAE) has been considered
in this work. It is shown that in this framework, fulfillment of the
TAE and properties 1. and 2. can only be matched with $G_{2}=0$.
Regarding these results it should be emphasized that, particularly
for the case of $G_{2}$ and the TAE, the models considered are tested
in the UV regime which is the regime where it is not at all clear
that they should work. Indeed, regarding these properties, the present
work can be considered as an exploration of the limitations of these
models in the UV. 

Interesting issues that deserve further study include: 
\begin{itemize}
\item The models considered in this work include the five dimensional metric
as a dynamical variable. A natural next step is to use the geometries
determined by these models as background metrics for 5-dimensional
holographic theories with fields corresponding to the basic observables
of QCD. The expectation is that the description of both IR and UV
properties should be improved in comparison with more crude assumptions\cite{Erlich:2005qh,Da_Rold:2005zs}.
\item Consideration of higher order in $1/\sqrt{N}$ corrections. This should
be considered as a means to study the stability of the results under
corrections. These corrections, from the point of view of a more fundamental
string theory motivating these background field actions, correspond
to corrections in $\alpha'$, the inverse of the string tension. They
correspond also to higher order curvature corrections and the more
fundamental string theory should provide precise expressions for the
higher curvature-higher dilaton derivatives terms. 
\end{itemize}
\bibliographystyle{plain}
\addcontentsline{toc}{section}{\refname}\bibliography{arxiv-pgqcd}

\end{document}